\documentclass{article}

\usepackage{epsfig}
\usepackage[T1]{fontenc}
\usepackage[latin1]{inputenc}
\usepackage[english]{babel}
\usepackage{times}
\usepackage{color}
\usepackage{subcaption}
\usepackage{amsthm,amsmath,enumerate}
\usepackage{amssymb}
\usepackage{amscd}
\usepackage{latexsym}
\usepackage{tabularx}
\usepackage{stmaryrd}
\usepackage{vmargin}
\usepackage{euscript}
\usepackage{xcolor}
\usepackage{a4}
\usepackage[colorlinks=true,citecolor=black,linkcolor=black,urlcolor=blue]{hyperref}

\setmarginsrb{3.5cm}{2cm}{3.5cm}{2cm}{1cm}{1cm}{1cm}{1cm}

\newtheorem{theorem}{Theorem}

\def\abs#1{\left| #1 \right|}
\def\paren#1{\left( #1 \right)}
\def\acc#1{\left\{ #1 \right\}}

\renewcommand{\le}{\leqslant}
\renewcommand{\ge}{\geqslant}

\title{$2$-subcoloring is NP-complete for planar comparability graphs}
\author{Pascal Ochem\\CNRS - LIRMM, Montpellier, France}

\begin{document}
\maketitle

\begin{abstract}
A $k$-subcoloring of a graph is a partition of the vertex set into at most $k$ cluster graphs, that is,
graphs with no induced $P_3$. $2$-subcoloring is known to be NP-complete for comparability graphs and three subclasses of
planar graphs, namely triangle-free planar graphs with maximum degree 4, planar perfect graphs with maximum degree 4,
and planar graphs with girth 5. We show that $2$-subcoloring is also NP-complete for planar comparability graphs with maximum degree 4.
\end{abstract}

%\begin{keyword}
%Graph coloring, NP-completeness.
%\end{keyword}

%\end{frontmatter}

\section{Introduction}
A $k$-subcoloring of a graph is a partition of the vertex set into at most $k$ cluster graphs, that is,
graphs with no induced $P_3$. Unlike $k$-coloring, $k$-subcoloring is already NP-complete for $k=2$:
\begin{theorem}
\label{known}
$2$-subcoloring is NP-complete for the following classes:
\begin{enumerate}[(1)]
 \item ($K_4$, bull, house, butterfly, gem, odd-hole)-free graphs with maximum degree 5~\cite{BFNW02}, \label{t1}
 \item triangle-free planar graphs with maximum degree 4~\cite{FKLS03,GH03}, \label{t2}
 \item ($K_{1,3},K_4,K_4^-,C_4$, odd-hole) planar graphs~\cite{GO}, \label{t3}
 \item planar graphs with girth 5~\cite{MO14}. \label{t4}
\end{enumerate}
\end{theorem}

A graph $G$ is \emph{$(d_1,\ldots,d_k)$-colorable} if the vertex set of $G$ can be partitioned
into subsets $V_1,\ldots,V_l$ such that the graph induced by the vertices of $V_i$
has maximum degree at most $d_i$ for every $1\le i\le k$.
Notice that every $(1,1)$-colorable graph is $2$-subcolorable.
Moreover, on triangle-free graphs, $(1,1)$-colorable is equivalent to $2$-subcolorable.
As it is well known, for every $a,b\ge0$, every graph with maximum degree $a+b+1$ is $(a,b)$-colorable~\cite{Lov66}.
Thus, every graph with maximum degree 3 is $2$-subcolorable, so that the degree bound of 4 in Theorems~\ref{known}.(\ref{t2}) and~\ref{known}.(\ref{t3}) is best possible.

Notice that the graphs in Theorem~\ref{known}.(\ref{t1}) are comparability graphs since they are (bull, house, odd-hole)-free~\cite{isgci}.
Our main result restricts the class in Theorem~\ref{known}.(\ref{t1}) to planar graphs and lowers the maximum degree from 5 to 4.
\begin{theorem}
\label{main}
Let $\mathcal{G}$ denote the class of ($K_4$, bull, house, butterfly, gem, odd-hole)-free planar graphs with maximum degree 4.
$2$-subcoloring is NP-complete for $\mathcal{G}$.
\end{theorem}

\section{Main result}\label{sec:hard}

\begin{figure}[htbp]
\begin{center}
\includegraphics[width=130mm]{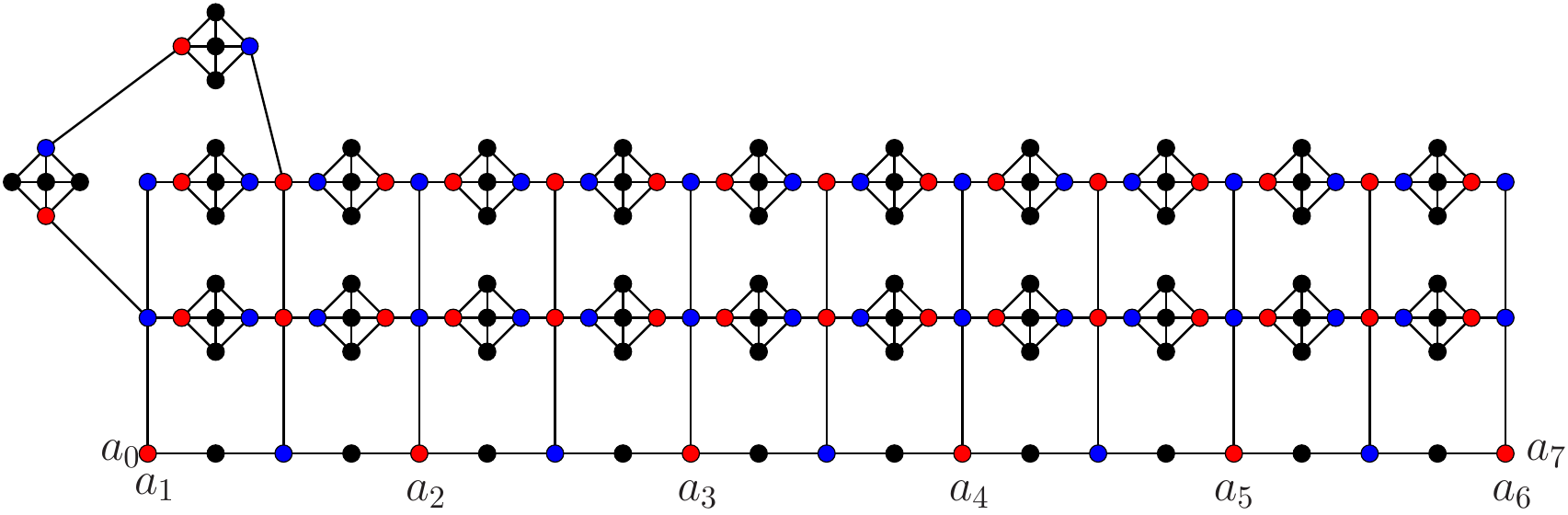}
\caption{The vertex gadget $H$.}
\label{vv}
\end{center}
\end{figure}

\begin{figure}[htbp]
\begin{center}
\includegraphics[width=60mm]{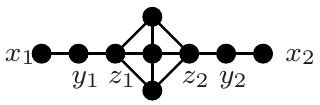}
\caption{The edge gadget $E$.}
\label{ee}
\end{center}
\end{figure}

The reduction is from $2$-subcoloring (or equivalently $(1,1)$-coloring) on triangle-free planar graphs with maximum degree 4,
which is NP-complete by Theorem~\ref{known}.(\ref{t2}).
From an instance graph $G$ of this problem, we construct a graph $G'$ in $\mathcal{G}$.
Every vertex $v$ of $G$ is replaced by a copy $H_v$ of the vertex gadget $H$ depicted in Figure~\ref{vv}.
For every edge $uv$ of $G$, we use two copies of the edge gadget $E$ depicted in Figure~\ref{ee} to connect $H_u$ and $H_v$ as follows:
\begin{itemize}
 \item We identify the vertex $x_1$ of the first (resp. second) edge gadget with a vertex $a_{2p}$ (resp. $a_{2p+1}$) of $H_u$, with $0\le p\le3$.
 \item We identify the vertex $x_2$ of the first (resp. second) edge gadget with a vertex $a_{i}$ (resp. $a_j$) of $H_v$
 such that $\min\paren{i,j}\equiv 0\pmod 2$, $\abs{j-i}=1$, and no edge crossing is created.
\end{itemize}
It is easy to check that $G'$ can be made planar and with maximum degree 4. Moreover, $G'$ is ($K_4$, bull, house, gem, butterfly)-free.
By removing the vertices whose neighborhood induces a $P_3$, we obtain a bipartite graph.
This shows that $G'$ is odd-hole free. Thus $G'$ belongs to $\mathcal{G}$.

Let us show that $G'$ is $2$-subcolorable if and only if $G$ is $2$-subcolorable.
Given a $2$-subcoloring of a graph, we say that a vertex $p$ is \emph{saturated} if there exists a monochromatic edge $pq$ and is unsaturated otherwise.
We will need the following properties of $E$:
\begin{enumerate}
 \item In every $2$-subcoloring of $E\setminus\acc{x_1,x_2,y_1,y_2}$, the vertices $z_1$ and $z_2$ get distinct colors and are saturated. \label{p1}
 \item In every $2$-subcoloring of $E\setminus\acc{x_1,x_2}$, the vertices $y_1$ and $y_2$ get distinct colors and are not saturated. \label{p2}
 \item There exists a $2$-subcoloring of $E$ such that the vertices $x_1$ and $x_2$ get distinct colors and are not saturated. \label{p3}
 \item In every $2$-subcoloring of $E$ such that the vertices $x_1$ and $x_2$ get the same color, exactly one vertex in $\acc{x_1,x_2}$ is saturated. \label{p4}
\end{enumerate}

The six vertices labeled $a_i$ in $H$ are called \emph{ports}.
Using properties (\ref{p1}) and (\ref{p2}), we obtain that every $2$-subcoloring of $H$ (in colors red and blue) forces the color of many vertices (see Figure~\ref{vv}).
In particular, all the ports in $H$ get the same color. This common color is said to be the color of $H_v$ corresponds to the color of $v$ in a $2$-subcoloring of $G$.
We also check that in every $2$-subcoloring of $H$, at most one of the ports is not saturated.

Suppose that $uv$ is an edge in $G$. Consider the $2$-subcolorings of the subgraph of $G'$ induced by $H_u$, $H_v$, and the two edge gadgets for the edge $uv$.
If distinct colors are given to $H_u$ and $H_v$, then this $2$-subcoloring can be extended to the edge gadgets using property (\ref{p3}).
Since this extension does not saturate any of the considered ports of $H_u$ and $H_v$,
$H_u$ can be connected to any number of vertex gadgets with the color distinct from the color of $H_u$.
If the same color is given to $H_u$ and $H_v$, then this $2$-subcoloring can be extended using property (\ref{p4}).
However, this coloring extension saturates the unique unsaturated port in both $H_u$ and $H_v$.
Thus, $H_u$ can be connected to at most one vertex gadget with the same color as $H_u$.

This shows that $G'$ is $2$-subcolorable if and only if $G$ is $2$-subcolorable.

\end{document}